# A FRAMEWORK FOR SOFTWARE-AS-A-SERVICE SELECTION AND PROVISIONING

Elarbi Badidi

College of Information Technology,
United Arab Emirates University, Al-Ain, United Arab Emirates

`ebadidi@uaeu.ac.ae`

## ABSTRACT

*As cloud computing is increasingly transforming the information technology landscape, organizations and businesses are exhibiting strong interest in Software-as-a-Service (SaaS) offerings that can help them increase business agility and reduce their operational costs. They increasingly demand services that can meet their functional and non-functional requirements. Given the plethora and the variety of SaaS offerings, we propose, in this paper, a framework for SaaS provisioning, which relies on brokered Service Level agreements (SLAs), between service consumers and SaaS providers. The Cloud Service Broker (CSB) helps service consumers find the right SaaS providers that can fulfil their functional and non-functional requirements. The proposed selection algorithm ranks potential SaaS providers by matching their offerings against the requirements of the service consumer using an aggregate utility function. Furthermore, the CSB is in charge of conducting SLA negotiation with selected SaaS providers, on behalf of service consumers, and performing SLA compliance monitoring.*

## 1. INTRODUCTION

Cloud computing permits a service-provisioning model, which typically involves the provisioning over the Internet of dynamically scalable and virtualized services. Applications or services offered by means of cloud computing are called *cloud services*. The three main models of cloud services are: *Infrastructure-as-a-Service (IaaS), Platform-as-a-Service (PaaS), and Software-as-a-Service (SaaS)*[1]

The advent of cloud computing has led to many repercussions on end-users and organizations. For end-users, a user using the cloud through a web-based application can access her documents and files whenever she wants and wherever she is, rather than having to remain at her desk. She can store more data in the cloud than on private computer systems. Documents and files stored in the cloud permanently exist, no matter what happens to the user's computer systems. In addition to this, cloud computing opens the door to group collaboration as the users from different physical locations can collaborate by sharing documents and files efficiently and at lower costs.For organizations, in addition to the above benefits, small and medium sized businesses can immediately take advantage of the enormous infrastructure of the cloud without having to implement and administer it directly. They can access multiple data centers from anywhere on the globe. This means that they can store more data than on their premises computer systems. It also means that their computing personnel no longer need to worry about keeping software up-to-date, but they will be free to concentrate more on innovation. If they need more processing power and more storage, it is always available in the cloud and accessible in a cost effective manner.

We believe that the success of cloud-based infrastructures will greatly depend on the level of satisfaction of customers in terms of performance and quality-of service (QoS) they get from cloud services providers. QoS refers to a collection of qualities or characteristics of a service, such as *availability*, *security*, *response-time*, *throughput*, *latency*, *reliability*, and *reputation* for the SaaS provisioning model. Such qualities are of interest to service providers and service consumers alike. They are of interest to service providers when implementing multiple service levels and priority-based admission mechanisms. The agreement between the customer and the service provider is referred to as the *Service Level Agreement* (SLA). An SLA describes agreed service functionality, cost, and qualities [2]. It is an agreement regarding the guarantees of the service provided by the cloud service provider. It defines mutual understandings and expectations of a service between the cloud service provider and service consumers. As of today, SLA-based service provisioning is being used in IaaS and PaaS service delivery models. However, its adoption in the SaaS service delivery model is still at its beginning. Cloud service providers typically use a resources' overprovision policy in their attempt to meet the SLA requirements from various customers, in terms of availability and performance [3]. Resources are statistically allocated to customers based on their requirements in the worst-case scenario. As a result of this policy, cloud provider servers may be sub-optimally used, as numerous allocated resources may be idle at run-time.

As the number of service customers is expected to grow in the coming years, it is vital for cloud service providers to overcome this situation by being able to allocate resources on an as-needed basis. Therefore, they should have the means to monitor resources' usage and evaluate various QoS metrics to be able to honor customers' SLA requirements. Given the limited number of solutions for SLA-based SaaS selection and provisioning and the heterogeneity of the APIs provided by each SaaS provider, we set out to develop a framework that will:

- Allow service consumers to express their functional and nonfunctional requirements.
- Hide the heterogeneity of SaaS providers' APIs from service consumers.
- Allow implementing QoS-driven selection of SaaS providers.
- Allow implementing SLA negotiation on behalf of the service consumer.
- Permit monitoring and assessment of SLAs execution.

The main component of the proposed framework is the *Cloud Service Broker (CSB)*, which mediates between service consumers and SaaS providers in order to reach SLAs, and implements the above management operations.

The remainder of this paper is organized as follows. The next section describes related work. Section 3 briefly describes background information on the concepts of cloud services and cloud service brokerage. Section 4 highlights the issues that organizations and businesses need to consider for the selection of potential SaaS offerings. Section 5 presents an overview of the proposed framework and describes the activities of its components. Section 6 describes the proposed algorithm for SaaS offerings' selection. Finally, Section 7 concludes the paper and describes future work.

## 2. RELATED WORK

Cloud service brokerage and the issues of SLA management, SLA negotiation in particular, are the subject of several research efforts over the last few years.
In the European project mOSAIC (www.mosaic-cloud.eu), an essential component is the Cloud Agency, which aims to allow cloud service consumers to delegate to the agency all SLA management tasks, the monitoring of resource utilization, and in some circumstances re-negotiation of SLA terms as the requirements of consumers may change [4]. The Cloud Agency system comprises several agents that collaborate to manage cloud resources and services offered by diverse cloud providers.

The SLA@SOI project is an ambitious project whose goal is defining a comprehensive view for the management of SLAs and developing a framework for SLA management that a service-oriented infrastructure (SOI) can incorporate[5]. Within this project, Theilmann et al. developed a reference architecture for multi-Level SLA management [6]. This architecture aims to offer a generic solution for SLA management that can: (1) support SLA management across multiple layers of a service-oriented infrastructure; (2) cover the complete SLA and service life cycle; and (3) be used in various industrial domains and use cases.

OPTIMIS [7] is a toolkit, which aims at optimizing the whole cloud service life cycle, including service creation, deployment, configuration, and operation, by considering non-functional issues such as *trust, risk, eco-efficiency and cost*. The toolkit targets mainly service providers and infrastructure providers and may also impact other actors such as brokers, and service consumers. The fundamental process during deployment is the negotiation of SLA terms between service providers and infrastructure providers.

Only few works, however, have investigated the issue of cloud services selection. Garg et al. [8] proposed the SMICloud framework for comparing and ranking cloud services by defining a set of attributes for the comparison of mainly IaaS cloud offerings. The ranking mechanism relies on the Analytical Hierarchical Process (AHP), which allows assigning weights to attributes in view of the interdependencies among them.Wang et al. [9] proposed a cloud model for the selection of Web services. This model relies on computing what the authors called *QoS uncertainty* and identifies the most appropriate Web services using mixed integer programming.Hussain et al. [10] proposed a multi-criteria decision making methodology for the selection of cloud services. To rank services, they match the user requirements against each service offering for each criterion. They also use the weighted difference or the exponential weighted difference methods to decide on the most suitable service.Similarly, Li et al. [11][12][13] developed *CloudCmp* a promising system for comparing offers from cloud providers in terms of performance and cost. CloudCmp uses ten metrics to compare the common services of cloud providers, mainly elastic computing cluster, persistent storage, intra-cloud and wide area network services. Results of the system will enable cloud users to predict the performance and cost of their applications before they deploy them on the cloud.Rehman et al. [14] proposed a system that relies on information collected by existing cloud users to make selection of cloud offerings. We think that this approach is not feasible in practice as many users will be reluctant to divulge information about the usage of their applications deployed in the cloud.

Our work shares with some of these efforts the common goal of mediating between service consumers and cloud service providers and providing support for automated SLA negotiation and management. We are focusing as a first step on the SaaS provisioning model. Finding the right SaaS provider is not an easy task for service consumers given the abundance and the variety of service offerings. Dealing with a SaaS provider requires knowledge of its operating environment, the availability of management tools, its security levels and data recovery approaches, and the service terms and conditions. Collecting this information for multiple service providers is likely to be a fastidious task that is costly and time consuming. The CSB with its know-how and value-added services will assist service consumers in: (a) finding appropriate SaaS offerings, (b) negotiating SLA terms, and (c) monitoring and assessing the implementation of SLAs.

## 3. BACKGROUND

### 3.1. Cloud Services

Applications or services offered by means of cloud computing are called *cloud services*. Typical examples of cloud services include office applications (word processing, spreadsheets, and presentations) that are traditionally desktop applications. Google Docs and Microsoft office Web Apps are cloud services of this category. Other examples include storage services, calendar, and notebooks' applications and many more. Many of these cloud-based applications are now available in the Chrome Web Store. Nearly, all large software corporations, such as Google, Microsoft, Amazon, IBM, and Oracle, are providing various kinds of cloud services. Besides, many small businesses have launched their own Web-based services, mainly to take advantage of the collaborative nature of cloud services.

The user of a cloud service has access to the service through a Web interface or via an API. Once started, the cloud service application acts as if it is a normal desktop application. The difference is that working documents are on the cloud servers. Cloud services offer all the benefits we have already mentioned. Cloud computing today is witnessing considerable interest from both academia and the computing industry. Many leading computing companies, such as Google, Amazon, Oracle, and Microsoft, are devoting resources for promoting cloud computing through the development of cloud services development tools. As we mentioned earlier in the introduction, Cloud services models are IaaS, PaaS, and SaaS.

One of the underlying advantages of the deployment of services in the cloud is the economy of scale. By making the most of the cloud infrastructure provided by a cloud vendor, a service provider can offer better, cheaper, and more reliable services than is possible within its premises. The cloud service can utilize the full processing and storage resources of the cloud infrastructure if needed. Another advantage is scalability in terms of computing resources. Service providers can scale up when additional resources are required as a result of a rise in the demand for their services. Conversely, they can scale down when the demand for service is decreasing. Another benefit of the approach is that it enables clients getting service on a pay-as-you-go basis and selecting cloud services based on the price and other criteria such as QoS. The net benefit for consumers and mobile users, in particular, is the ability to receive better services tailored to their current needs.

SaaS represents the trend of the future and the most common form of cloud service development. With SaaS, software is deployed over the Internet and delivered to thousands of customers. Using this model, the cloud service provider may license its service to customers through a subscription or a pay-as-you-go model. The service is then accessible using an API.

### 3.2. Cloud Service Brokerage

As cloud computing technology matures, cloud services offers are proliferating at an unprecedented pace. As in every business with a delivery model, such as real estate and insurance, cloud services' brokerage will emerge in order to enable organizations to procure cloud services efficiently. Indeed, finding the right cloud service is not an easy task for service consumers given the plethora and the variety of cloud services offerings. Dealing with a cloud service provider requires knowledge of its operating environment, the availability of management tools, its security levels and data recovery approaches, and the service terms and conditions. Collecting this information for multiple cloud service providers is likely to be a demanding task that is expensive and time consuming. Cloud service brokers (CSBs) with their know-how and

International Journal of Computer Networks & Communications (IJCNC) Vol.5, No.3, May 2013

value-added services will assist service consumers in finding appropriate cloud service offerings, carrying out the SLA negotiation process, monitoring and assessing the implementation of SLAs. Gartner predicts that in parallel with the growing adoption of cloud services CSBs will emerge. They will mainly be in charge of the management of the utilization, performance, and delivery of cloud services. CSBs will broker relationships between service consumer and multiple cloud providers [8].

*"The future of cloud computing will be permeated with the notion of brokers negotiating relationships between providers of cloud services and the service customers. In this context, a broker might be software, appliances, platforms or suites of technologies that enhance the base services available through the cloud. Enhancement will include managing access to these services, providing greater security or even creating completely new services,"* [15].

The NIST identified in its Cloud Computing Reference Model the Cloud Broker actor, which is in charge of service intermediation, service aggregation, and service arbitrage [16].

The concept of brokerage service is not new in IT; telecommunication and traditional distributed systems have used brokers in order to manage quality-of-service. Recently, several SOA-based systems have used brokers to mediate between clients and service providers [17][18][19]. The broker's goal is to facilitate the transactions between service requesters and providers, create a seamlessly trusted environment, and make recommendations as accurate as possible. The functions of the service broker typically include monitoring and collecting QoS information of Web services, making selection decisions on-behalf of clients, and negotiating SLAs and QoS assurances with Web services.

Moore et al. [20] designed and implemented a Web service broker to help manage the interactions and data exchange between clients looking up for services and SaaS providers interested in promoting their services in a trusted environment. The service broker acts as a container for publishing heterogeneous SaaS applications from various SaaS providers. It manages trust issues between the provider and the client, and thus allows providers to lower their own trust policies.

## 4. ISSUES TO CONSIDER FOR SAAS PROVIDERS' SELECTION

Organizations and consumers who choose to use SaaS services need to consider several issues before relying on SaaS provisioning. These issues concern essentially service functionality, integration with on-premises services, SLA negotiation and compliance monitoring, software change management, data access, and data security.

*Service functionality* - The organization has to make sure that the SaaS offering has almost all the features of the on-site, internally deployed service in the case of an existing on-premises service and all the necessary features in the case of a new service that will integrate with existing services. Moreover, the organization should check whether the SaaS offering is about a differentiated service with, for example, platinum, gold, and silver service levels.

*Integration with on-premises services* – The organization should be aware of any requirements for the integration of the SaaS offering with its existing, internal, on-site applications. For instance, the conformity of the data format used internally with the format used by the SaaS offering and network security requirements for the integration should be assessed.

*SLA negotiation and compliance monitoring* - The organization has to make sure that: (1) an SLA governs the service provisioning; (2) the two parties need to negotiate the terms of the SLA; (3) it has the necessary information on the SLA options available from the SaaS offering; (4) the

International Journal of Computer Networks & Communications (IJCNC) Vol.5, No.3, May 2013

SLA meets its business needs; (5) the SaaS provider monitors its service provisioning and evaluates its compliance with the SLA; (6) it can rely on third parties, such as keynote(keynote.com), for SLA compliance monitoring; and (7) the SLA includes penalty and compensation sections for SLA noncompliance.

*Software change management* – Software change management is part of any SaaS offering as new versions of software are developed and deployed to meet emerging business needs. Since the need to document, track and manage software doesn't change, the organization needs to realize in advance whether: (1) the SaaS provider will handle patches, service level releases and other upgrades consistently with its expectations; (2) the SaaS provider will carry out any change to the SaaS offering's environment in a similar test environment before its promotion to production; and (3) the organization will contribute to any upgrade and testing of new versions of the software.

*Data storage and access* - Data storage and access are critical issues for enterprises. Therefore, it is extremely vital for the organization to know in advance: (1) how the SaaS provider is going to use the organization data; (2) whether the SaaS provider's privacy policy is consistent with the expectations of the organization regarding the utilization of its data; (3) whether the SaaS provider allows importing from and exporting data to the SaaS solution; (4) whether it has full access to its data within the SaaS offering; (5) whether its data is totally isolated from any other client of the SaaS offering; and (6) whether the SaaS provider will destroy from its storage the organization's data if it terminates the agreement with the SaaS provider.

*Data security* – Data security is also a critical matter for the organization. Indeed, it is crucial for the organization to know in advance whether: (1) the SaaS provider has in place anti-theft mechanisms; (2) it conducts security tests and third party audit on a regular basis; (3) it offers legal commitments concerning their security measures; and (4) the SaaS provider's country offers necessary legal protection of the organization's data.

## 5. FRAMEWORK OVERVIEW

Figure 1 depicts our framework for SLA-based service provisioning. The main components of the framework are: *Service consumers (SCs)*, *Cloud Service Broker (CSB)*, *Measurement Services*, and *SaaS Providers (CSPs)*.

As we mentioned earlier, the Cloud Service Broker (CSB) is a mediator service that decouples service consumers from SaaS providers. It is in charge of handling subscriptions of service consumers in which they express their interest to obtain some kinds of service, and registration of SaaS providers that are willing to provide some types of service. Given that service consumers do not normally have the capabilities to negotiate, manage, and monitor QoS, they delegate management tasks, such as SaaS providers' selection and SLA negotiation, to the Cloud Service Broker. The Cloud Service Broker can be typically used in the following scenario: A University has a number of graduate students and Colleges' employees that want to use an external public cloud service, e.g. a grammar checker or bibliographic references and abstracts provider. The CSB acts as a secure gateway that translates the Ids of the students and colleges' staff to the credentials used to access the external cloud service. It actively monitors the implementation of the SLA and generates reports on the service usage by each college.

The Cloud Service Broker performs several management operations to deliver personalized services to consumers. These operations are: *Identity and Access Management* (IAM), *Policies Management* (PM), *SLA Management* (SLAM), and *Service Provisioning* (SP). They are under the control of a *Coordinator* component. The back-end databases maintain information about services' policies, consumers' profiles and preferences, SLAs, and dynamic QoS information.

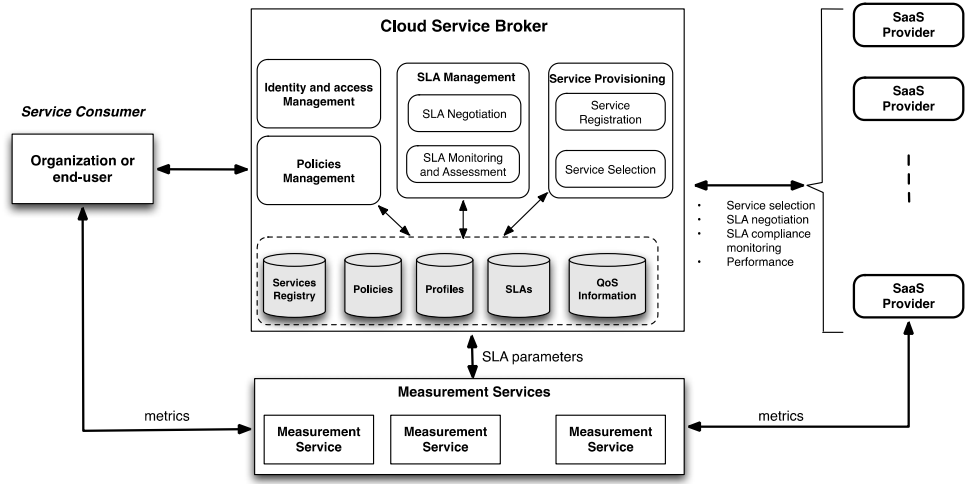

Figure 1. Framework for SaaS selection and provisioning

Several competing SaaS providers may provide the same service type. Therefore, potential service consumers should be able to select SaaS providers on the basis of the QoS they can assure and on their commitment to negotiated SLA. The *Selection Manager* that implements SP management operations is in charge of implementing different policies for the selection of suitable SaaS providers, based on the consumer's QoS requirements and the SaaS providers' QoS offerings.

The *SLA Manager*, which implements SLAM management operations, is in charge of carrying out the negotiation process between a consumer and a selected SaaS provider in order to reach an agreement as to the service terms and conditions. It approaches this SaaS provider to determine whether it can ensure the required level of service given its current conditions. Then, the consumer and the SaaS provider sign a contract. The contract specifies the service that the provider should offer, the level of service to ensure, the cost of service, and actions to take when several violations of the agreement occur. If the selected SaaS provider is unable to deliver the required level of service, the CSB selects another SaaS provider and reiterates the negotiation process.

The *Profile Manager*, which implements IAM management operations, is responsible for managing clients' profiles, including their preferences in terms of personalized services and required QoS. The *Policy Manager, which implements PM management operations,* is responsible for managing different kinds of policies such as authorization policies and QoS-aware selection policies of service providers.

SaaS providers typically offer several types of services using Web services. These Web services can be simple or composite Web services. Composite Web services are the result of the composition of many simple or composite services. In order to estimate their current QoS for each service type they offer, service providers should use monitoring techniques to collect measurement data at selected observation points. By aggregating collected data, the SaaS provider can determine the value of each QoS indicator and checks its compliance with the SLA. If there is a significant drop in the actual QoS offering, the SaaS provider might take appropriate actions such as adding more resources to avoid any complaint from the service consumer. Figure 4 shows a typical architecture of a SaaS Provider, which includes several management functions that cooperate in order to deliver personalized services to consumers. These functions are: *Identity and Access Management* (IAM), *Policies Management* (PM), *SLA Management* (SLAM),

*Service Provisioning* (SP), and *Measurements and QoS Evaluation (MQE)*. The back-end databases maintain information about services' policies, consumers' profiles, SLA templates, Committed SLAs, and dynamic QoS information.The *Measurements and QoS Evaluation* subsystem collects data on QoS indicators at various observation points and estimates current QoS offerings. The *SLA Manager* component is responsible for negotiating with the Cloud Service Broker, or directly with service consumers, the terms of service and the QoS level to be delivered.

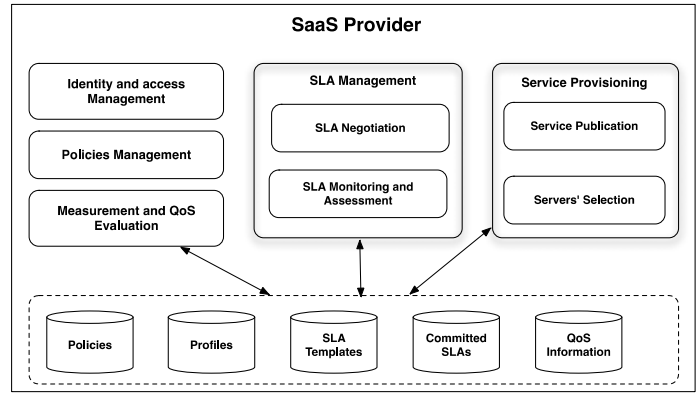

Figure 2. Typical architecture of a SaaS provider infrastructure

*Measurement services*, as depicted in figure 1, are third parties actors that have become new players in the cloud ecosystem as they allow service providers to have independent monitoring of their offerings and allow service consumers to make sure that they are getting the level of service that the service provider promised. Examples of measurement and monitoring services are keynote (keynote.com), Monitis (monitis.com), and Uptrends (uptrends.com). Measurement services are in charge of mapping resource metrics into SLA parameters and monitoring current service levels and their compliance with SLA. Plugins to these measurement services may be added to the framework as service consumers and SaaS providers might prefer specific monitoring companies.

## 6. A SaaS Selection Algorithm

As we mentioned earlier in the related work section, a limited number of works have investigated the issue of cloud services selection. In this section, we describe an algorithm for the selection of SaaS providers that can fulfill the service consumer request. The algorithm considers only non-functional (mainly QoS) issues in the selection process. The other issues we discussed in section 3 are negotiated during the SLA negotiation process once the CSB has selected a potential SaaS provider.

A service consumer may have her own preferences in terms of QoS, for the service she would like to obtain. Similarly, SaaS providers may have different QoS offerings for each service they are providing. We assume that QoS indicators are in normalized form with values between 0 and 1. A value of 1 means highest quality and 0 means lowest quality. The Cloud Service Broker, upon reception of a SLA request, maps the service consumer expectations into normalized QoS values.

Let $P = \{X_1, X_2, \ldots, X_n\}$ be the vector of QoS parameters considered in the system. Let $M = \{m_1, m_2, \ldots, m_n\}$, with $0 \leq min_i \leq 1$, be the vector of minimum quality requirements that the service consumer tolerates for the required type of service. These values are obtained by mapping the quality expectations (platinum, gold, silver) of the service consumer into normalized form. Let $S_i = \{SP_1, SP_2, ..., SP_k\}$ be the set of SaaS providers that can provide the service requested by the service consumer. Let $Q_i = \{q_{1i}, q_{2i}, ..., q_{ni}\}$, with $0 \leq q_{ji} \leq 1$, be the QoS offering of the SaaS provider $SP_i$.

The CSB's Selection Manager evaluates an aggregate utility function and determines whether the offer of *a SaaS provider* is acceptable or not. If we assume that the QoS attributes are independent, the linear aggregate utility function can be defined by:

$$U = w_1 U_1 + w_2 U_2 + \cdots w_m U_m \tag{1}$$

with $\sum_1^m w_i = 1$.

$U_i$ represents the individual utility function associated with the QoS attribute $X_i$, and $w_i$ is the weight that the service consumer assigns to that attribute. The service provider offer is acceptable to the CSB if the aggregate utility function exceeds a predefined threshold, which is calculated from the service consumer expectations.

Various functions may be used to express the service consumer utility of an attribute $X_i$. We borrow the function used in [21] to express that utility function as:

$$U_i = x_i^{\beta_i} \tag{2}$$

$\beta_i$ is a measure of the service consumer sensitivity to the QoS attribute $x_i$. When $\beta_i = 0$, the service consumer is indifferent to QoS attribute $x_i$. When $\beta_i = 1$, the service consumer is moderately sensitive to QoS attribute $x_i$ (the relationship is linear). When $\beta_i > 1$, the service consumer is increasingly sensitive to QoS attribute $x_i$. As $\beta_i$ increases, the service consumer is expressing increasing concern about $x_i$. For $\beta_i < 1$, as $\beta_i$ decreases to approach 0, the service consumer is expressing increasing indifference to having $x_i$.

Equation (1) becomes:

$$U = w_1 x_1^{\beta_1} + w_2 x_2^{\beta_2} + \cdots w_n x_n^{\beta_n} \tag{3}$$

As a proof of concept, we consider a scenario where the QoS parameters considered in the system are: *availability*, *1/response-time*, *reliability*, and *throughput*. Table 1 depicts normalized minimum QoS requirements of the service consumer, the weight and the sensitivity value associated to each quality parameter.

Table1.Minimum QoS requirements of the service consumer

|  | **Availability** | **1/RT** | **Reliability** | **Throughput** |
|---|---|---|---|---|
| $\omega_i$ | 0.35 | 0.15 | 0.35 | 0.15 |
| $m_i$ | 0.98 | 0.65 | 0.95 | 0.90 |
| $\beta_i$ | 1 | 1 | 1 | 1 |

Assume that four SaaS providers, as shown in Table 2, have submitted their QoS offerings to the CSB. Theses providers are: $SP_1$, $SP_2$, $SP_3$, and $SP_4$.

Table2. QoS offering of four potential SaaS providers

|  | Availability | 1/RT | Reliability | Throughput |
|---|---|---|---|---|
| $SP_1$ | 0.94 | 0.70 | 0.98 | 0.70 |
| $SP_2$ | 0.98 | 0.60 | 0.97 | 0.65 |
| $SP_3$ | 0.97 | 0.80 | 0.96 | 0.75 |
| $SP_4$ | 0.98 | 0.85 | 0.98 | 0.70 |

By computing the aggregate utility function for each SaaS provider, we get the following ranking from highest offer to lowest:

$$SP_4\,(0.92),\ SP_3\,(0.91),\ SP_1\,(0.88),\ SP_2\,(0.87)$$

The value between parentheses corresponds to the computed aggregate utility. The utility value of the service consumer computed from the minimum quality requirements is 0.91. This means that the QoS offerings of $SP_1$ and $SP_2$ do not meet the minimum QoS requirements of the service consumer.

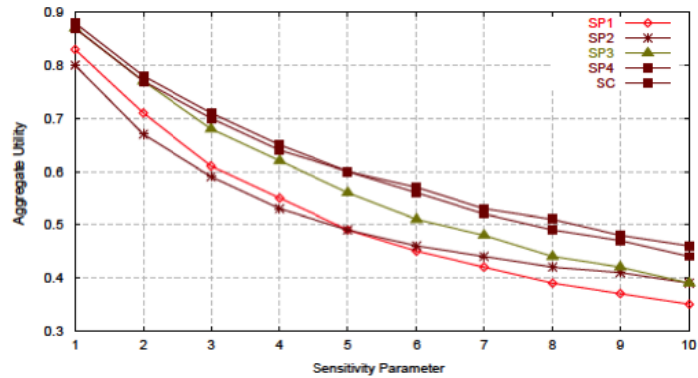

Figure 3. Evolution of the aggregate utility with the sensitivity parameter

Figure 3 shows the variation in the aggregate utility function of the service consumer and the four SaaS offerings as a function of the sensitivity parameter. To make the scenario simple, the sensitivity parameter is the same for all quality parameters. In this scenario, the SaaS provider $SP_4$ has the most promising offering as its aggregate utility exceeds the expectations of the service consumer. $SP_3$ offering is very close to these expectations. $SP_2$ and $SP_1$ offerings are below these expectations.

## 7. CONCLUSION AND FUTURE WORK

Given the wide adoption of cloud computing technology and the growing number of SaaS providers, service consumers will increasingly face the challenge of finding appropriate SaaS providers that can satisfy their functional and non-functional requirements.

In this paper, we have presented a framework for SaaS selection and provisioning. The framework relies on a Cloud Service Broker, which is in charge of mediating between service consumers and SaaS providers and negotiating the SLA terms. The proposed SaaS providers' selection algorithm uses a linear aggregate utility function, which assumes that the various QoS parameters are independent, to rank the potential SaaS offerings by matching them against the quality requirements of the service consumer. We describe in another article the interfaces of the various components of the Cloud Service Broker, the SLA negotiation process, and SLA compliance monitoring. As a future work, we intend to define a common ontology for QoS representation and the mapping from the diverse representations used by SaaS providers to that common ontology; and build a prototype of the framework together with some real scenarios for SLA-based service selection and provisioning.

**Authors**

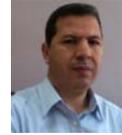

**Dr. Elarbi Badidi** received a Ph.D. degree in computer science in 2000 from Université de Montréal, Québec (Canada). He joined the College of Information Technology (CIT) atthe UAE University in Fall 2004 where he is currently an associate professor of computer science. Before joining the CIT, Dr. Badidi worked at the ENSIAS school of engineering in Rabat (Morocco) and Université de Montréal. Dr. Badidi has been conducting research in the areas of object-based distributed systems, bioinformatics, and Web services. He served on the technical program committees of many international conferences. His research interests include Web services and Service Oriented Architectures, cloud computing, context-aware services, bioinformatics, and middleware.